# pPython for Parallel Python Programming

Chansup Byun, William Arcand, David Bestor, Bill Bergeron, Vijay Gadepally, Michael Houle, Matthew Hubbell,
Hayden Jananthan, Michael Jones, Kurt Keville, Anna Klein, Peter Michaleas, Lauren Milechin, Guillermo Morales,
Julie Mullen, Andrew Prout, Albert Reuther, Antonio Rosa, Siddharth Samsi, Charles Yee, Jeremy Kepner
Massachusetts Institute of Technology

*Abstract*—pPython seeks to provide a parallel capability that provides good speed-up without sacrificing the ease of programming in Python by implementing partitioned global array semantics (PGAS) on top of a simple file-based messaging library (PythonMPI) in pure Python. The core data structure in pPython is a distributed numerical array whose distribution onto multiple processors is specified with a 'map' construct. Communication operations between distributed arrays are abstracted away from the user and pPython transparently supports redistribution between any block-cyclic-overlapped distributions in up to four dimensions. pPython follows a SPMD (single program multiple data) model of computation. pPython runs on any combination of heterogeneous systems that support Python, including Windows, Linux, and MacOS operating systems. In addition to running transparently on single-node (e.g., a laptop), pPython provides a scheduler interface, so that pPython can be executed in a massively parallel computing environment. The initial implementation uses the Slurm scheduler. Performance of pPython on the HPC Challenge benchmark suite demonstrates both ease of programming and scalability.

*Keywords—PGAS, Python, MPI, Parallel Programming*

I. INTRODUCTION

Python is one of the most widely used programming languages among developers around the world [1]. Python has become popular among scientific and engineering computing communities because Python is open source, its syntax is easy to understand, and rich ecosystem of scientific and mathematical packages has developed, such as, NumPy [2] and SciPy [3]. Furthermore, many different approaches have been developed for parallel programming with Python on shared and distributed memory environments. A comprehensive list of parallel Python libraries and software is available in Ref. 4.

Among the various approaches cited in Ref. 4, the message passing approach [5-10] appears to be one of the most widely used approaches. The message passing approach requires the user to explicitly send messages within the code. These approaches often implement a variant of the Message Passing Interface (MPI) standard [11]. Message passing allows any processor to directly communicate with any other processor and provides the minimum required functionality to implement a parallel program. Users that are already familiar with MPI find these approaches powerful. However, the learning curve is steep for the typical user because explicit message passing approaches significantly lower the level of abstraction and require users to deal directly with deadlocks, synchronization, and other low level parallel programming concepts. In addition, the impact on code size is significant. Serial programs converted to parallel programs with MPI typically increase in size by 25% to 50%; in contrast, OpenMP and PGAS approaches typically only increase the code size by ~5% [12].

In spite of these difficulties, a message passing capability is a requirement for other parallel programming approaches such as client-server or global arrays for a distributed memory programming. Furthermore, message passing is often the most efficient way to implement a program and there are certain programs with complex communication patterns that can only be implemented with direct message passing. Thus, any complete parallel solution must provide a mechanism for accessing the underlying messaging layer. Among the available Python message passing implementations, mpi4py [5] is widely used [13] and actively maintained.

Although a number of other parallelization approaches have been developed for Python [4], there are a limited number of published works on the parallelization of Python with global arrays. The extensive list of work related to global arrays is surveyed in Ref. 14. In particular, there have been a couple of publications on implementing global arrays in Python [13,15].

The MIT Lincoln Laboratory Supercomputing Center (LLSC) has focused on developing a unique, interactive, on-demand high-performance computing (HPC) environment to support diverse science and engineering applications. This system architecture has evolved into the MIT SuperCloud. MIT SuperCloud not only continues to support parallel MATLAB and Octave jobs, but also jobs in Python [16], Julia [17], R [18], TensorFlow [19], PyTorch [20], and Caffe [21] along with parallel C, C++, Fortran, and Java applications with various flavors of message passing interface (MPI) [11].

One of the core software stacks at LLSC environment is pMatlab [22,23], which implements Partitioned Global Array Semantics (PGAS) [24] using standard operator overloading techniques. pMatlab includes MatlabMPI [25], which provides MPI point-to-point communication and the gridMatlab [26] scheduler interface. pMatlab has subsequently inspired the MathWorks parallel computing toolbox used by many thousands of scientists and engineers around the world. pPython seeks to provide all the benefits available with pMatlab, MatlabMPI, and gridMatlab in a Python programming environment.

The core data structure in pPython is a distributed numerical array whose distribution onto multiple processors is specified with a 'map' construct. Communication operations between

This material is based upon work supported by the Assistant Secretary of Defense for Research and Engineering under Air Force Contract No. FA8721-05-C-0002 and/or FA8702-15-D-0001. Any opinions, findings, conclusions or recommendations expressed in this material are those of the author(s) and do not necessarily reflect the views of the Assistant Secretary of Defense for Research and Engineering.

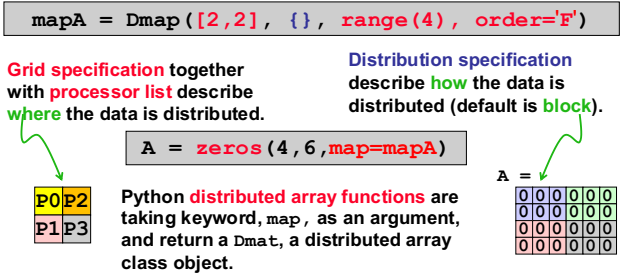

Fig. 1. Anatomy of a map. A map for a numerical array is an assignment of blocks of data to processing elements. It consists of a grid specification (in this case, {} implies that the default block distribution should be used), and a processor list (in this case the array is mapped to processors, 0., 1, 2, and 3). In pPython, grid specification can choose the ordering of processors in column or row direction with the keyword, order.

distributed arrays are abstracted away from the user and pPython transparently supports redistribution between any block-cyclic-overlapped distributions in up to four dimensions. pPython is built on top of the PythonMPI communication library and runs on any combination of heterogeneous systems that support Python, which includes Windows, Linux, MacOS operating systems. In addition, pPython includes a scheduler interface and enables users to submit their computing tasks via the scheduler.

In the subsequent description of pPython the widely used HPC Challenge parallel computing benchmarks STREAM, FFT, High Performance Linpack (HPL), and RandomAccess are used to illustrate pPython implementation and performance.

## II. pPython Interface and Architecture Design

pPython interface and architecture design are implemented in pure Python and to provide ease-of-use, high performance, and ease-of-implementation.

### A. Ease-of-Use

pPython adopts a separation-of-concerns approach to make program correctness and mapping a program to a parallel architecture orthogonal. A serial program is made parallel by adding maps to arrays. Maps only contain information about how an array is broken up onto multiple processors and the addition of a map does not change the functional correctness of a program. A map (Fig. 1) is composed of a grid specifying how each dimension is partitioned, a distribution that selects either a block, cyclic or block-cyclic partitioning, and a list of processor IDs that defines which processors actually hold the data.

In Python, when an array is created, its default ordering is row-major order (C-style). In order to provide flexibility with ordering the processor grid, the keyword 'order' is introduced with the map function. This allows to support a processor grid arranged in the column–major order (Fortran style). It is also noted that the map function name itself is also changed to 'Dmap' because Python already has a 'map' function for a different purpose. In pPython, when creating a distributed array as shown in Fig. 1, the keyword 'map' is introduced for a distributed array. Without the keyword 'map', the distributed array functions, zeros, ones, and rand, will return as standard NumPy arrays, which simplifies going from serial to parallel programs.

The next step in writing a parallel program is implementing communications. Redistribution between any two distributed arrays in pPython is accomplished with the __setitem__ method in the distributed array class, Dmat, by defining the subscripted assignment (subsasgn) operator ('=' with an indexed left-hand side). For example, in the STREAM benchmark (Fig. 2) the '=' operator was used in the statement: A[:,:] = B + s*C, but, since the arrays A, B and C all have the same map, no communication was required. The __setitem__ method in pPython determines when communication is required and correctly performs a simple assignment of the local data on the right-hand side to the local data on left hand side. A more complex example is the HPC Challenge FFT benchmark (Fig. 3). This benchmark computes the fast Fourier transform of a large 1-D vector. The standard parallel algorithm for this benchmark is to transform the 1-D vector into a row distributed matrix, FFT the rows of the matrix, multiply by a set of weights, redistribute into a column distributed matrix, and FFT the columns. A key step in the process is the redistribution which is performed by the statement: Z[:,:] = X, which determines and executes the Np$^2$ messages that need to be sent in order to complete this operation (where Np is the number of distinct pPython instances that are working in parallel). By default, the entire length is used when doing FFT in rows but 'len(Z)' is used to provide the index length for the column direction.

```
Np = pPython.Np          # Set number of processors.
N = 32                   # Set size of row vector.
S = 3.14                 # A scalar value.

ABCmap = Dmap([1,Np],{},range(Np)) # Create a map
A = zeros(1,N,map=ABCmap)  # Distributed zero array.
B = rand(1,N,map=ABCmap)   # Distributed random array.
C = rand(1,N,map=ABCmap)   # Distributed random array.

A[:,:] = B + s*C         # Local scale and add.
```

$$A = B + s*C$$

with blocks labeled $0, 1, \cdots, N_P-1$ for each of A, B, and $s*C$.

Fig. 2. STREAM benchmark code highlights. The first three lines set the various constants required by the program such as the number of processors and the size of the row vector. The next line creates a map, which will cause the second dimension of a distributed array to be broken up equally among all the processors. The next three lines use this map to create three row vectors. The last line performs the basic STREAM triad arithmetic operations in parallel. No communication is required in this example because A, B and C are all mapped the same.

```
Np = pPython.Np          # Set number of processors.
P = 2**10; Q. = 2**10    # Set dimensions of array.

Xmap = Dmap([Np,1],{},range(Np)) # Row map
Zmap = Dmap([1,Np],{},range(Np)) # Column map
# Create complex global arrays X and Z for FFT.
X = dcomplex(rand(P,Q,map=Xmap),rand(P,Q,map=Xmap))
Z = dcomplex(zeros(P,Q,map=Zmap),zeros(P,Q,map=Zmap))

X = fft(X,axis=1)    # FFT rows.
X = X * W            # Element-wise multiply by weights.
Z[:,:] = X           # Redistribute data.
Z = fft(Z,len(Z),0)  # FFT columns.
```

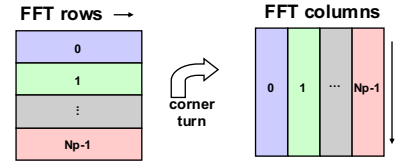

Fig. 3. FFT benchmark code highlights. The first two lines set the various constants required by the program such as the number of processors and the size of the matrix. The next two lines create two map objects for breaking the matrix up into rows and into columns. The next two lines use the maps to create two matrices. The next four lines FFT the rows, multiply by a set of local pre-computed weights, redistribute the data (using the "=" operator) into the matrix broken up by columns, and then perform the FFT on the columns.

PGAS enables complex data movements to be expressed compactly without making parallelism a burden to code. For example, removing the maps from either the STREAM or FFT examples returns the program to a valid serial program that simply uses standard built-in operations. This is a direct result of the orthogonality of mapping and functionality, and allows the pPython library to be "turned off" by simply setting all the maps equal to the scalar value of one. This feature is implemented in the Dmat class so that pPython distributed array functions (e.g., zeros and rand), return a NumPy array if map is not a Dmap class object. This ability to turn the library on and off is a key debugging feature and allows users to determine whether the bugs are from problems in their serial code or due to their use of pPython constructs.

*B. High Performance*

pPython adopts a coding style that uses some fragmented PGAS constructs (see Ref. 22 for definition of fragmented PGAS and details about the choice). This style is less elegant but provides strict guarantees on performance. More specifically, distributed arrays are used as little as possible and only when interprocessor communication is required.

However, PGAS constructs are not appropriate for all circumstances. There are communication patterns that would be more efficient if direct message passing can be employed. Thus, it is important to have mechanisms that allow PGAS and the underlying communication constructs to interact easily. pPython provides this ability by allowing the user to directly access the underlying PythonMPI library and its data structures. Several of the HPC Challenge benchmarks fall into the class of codes that do best by allowing some use of direct message passing [22]. HPC Challenge parallel computing benchmarks STREAM, FFT, High Performance Linpack (HPL), and RandomAccess are implemented with pPython to demonstrate this ability.

*C. Ease-of-Implementation*

Considering to achieve balance between the ease-of-use and high-performance [22], one of the key choices in implementing a PGAS library is which data distributions (see Section III.B) to support. At one extreme, it can be argued that most users are satisfied by 1-D block distributions. At the other extreme, one can find applications that require truly arbitrary distributions of array indices to processors. pPython has chosen to support up to 4-D block-cyclic distributions with overlap because the problem of redistribution between any two such distributions (see Section III.C) is highly complex to program for the use but has been solved a number of times by different parallel computing technologies. This allows pPython users to create four-dimensional arrays with all four dimensions distributed.

The subscripted assignment operator (such as A[i:j,k:l]=B) is implemented as the Python __setitem__ method, which supports data redistribution between arrays. The next question is, what other functions should we support and for which distributions? A challenge is overloading all mathematical functions in a library to work well with every combination of input distributions. This capability is extremely difficult to implement and is not entirely necessary if users are willing to tolerate slightly less elegant coding style associated with fragmented PGAS. Thus, pPython provides a rich set of data distributions, but a relatively modest set of overloaded functions, which are mainly focused on array construction functions, array index support functions, and the various element-wise operations (+, −, *, /, ...).

III. pPYTHON IMPLEMENTATION

This section discusses the implementation of the pPython library. pPython employs a layered architecture [22]. Additionally, pPython has added a scheduler interface for a grid environment such as the LLSC environment. In the layered architecture, the pPython library implements distributed constructs, such as distributed matrices and higher dimensional arrays. pPython also provides parallel implementations of a select number of functions such as redistribution, Fast Fourier Transform (FFT), and currently porting of matrix multiplication is in progress at the time of writing this paper.

The pPython library uses the parallelism through polymorphism approach as discussed by Choy and Edelman [27]. In addition, the polymorphism is further exploited by introducing the map class object. Map objects belong to a pPython Dmap class and are created by specifying the grid description, distribution description, and the processor list (Fig.

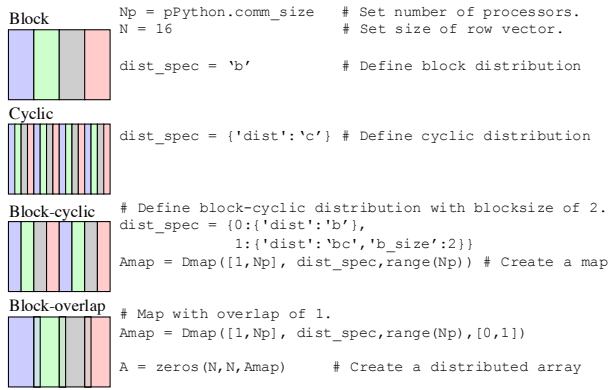

Fig. 4. Supported distributions. Block distribution divides the object evenly among available processors. Cyclic distribution places a single element on each available processor and then repeats. Block-cyclic distribution places the specified number of elements on each available processor and then repeats.

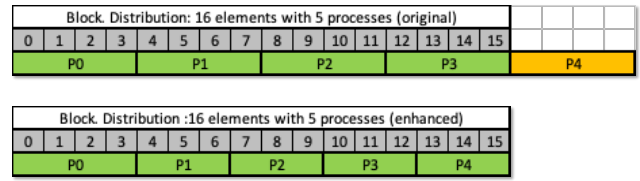

Fig. 5. PITFALLS implementation enhancement with the block distribution. All the processes will get their fair-share when the dimension size is not a multiple of the number of processers under a certain condition where no element will be assigned to one or more processers.

1). The map object can then be passed to a pPython method, such as `rand`, `zeros`, or `ones`. These methods are implemented in a way that, when a map object is passed, the library creates a distributed array class, `Dmat`, object. A PITFALLS [28] structure (Section III.C) is created when each `Dmat` object is constructed. A PITFALLS structure is a mathematical representation of the data distribution information. pPython supports numerical arrays of up to four dimensions of different numerical data types, including the support of distributed sparse matrices.

Since all functions supported in pPython are implemented in pure Python, pPython can run anywhere Python runs, given that there exists a common file system, a constraint imposed on pPython by the PythonMPI file-based messaging library. A further benefit of the layered architecture of pPython is that any other communication library could be substituted for PythonMPI (Section III.D).

*A. pPython Execution*

All pPython code resides within a generic code framework for determining the number of processors the program is being run on (`pPython.Np`) and determining the rank of the local processor (`pPython.Pid`). The initializing pPython (`pPython_init`) library is abstracted away from the user and is called when the `pRUN()` method is invoked.

pPython uses the single program multiple data (SPMD) execution model. The user runs a pPython program by executing the pPython pRUN command to launch and initialize the multiple instances of Python required to run in parallel.

*B. Maps and Distributions*

The pPython map construct is defined by three components: (1) grid description, (2) distribution description, and (3) processor list. The grid description together with the processor list describes where the data object is distributed, while the distribution describes how the object is distributed (Fig. 1). The syntax for defining these distributions in pPython is shown in Fig. 4.

Block distribution is the default distribution, which can be specified explicitly or by simply passing an empty distribution specification to the map constructor. Cyclic and block-cyclic distributions require the user to provide more information. Distributions can be defined for each dimension and each dimension could potentially have a different distribution scheme. Furthermore, distributions can be a partial set of processors which naturally enables distributed streaming applications. Additionally, if only a single distribution is specified and the grid indicates that more than one dimension is distributed, that distribution is applied to each dimension. Distribution specification is described by a Python dictionary variable for a general specification but it can be defined by a string for block, 'b', or cyclic, 'c', distribution when only a single distribution is specified for all dimensions.

Some applications, particularly image processing, require data overlap, or replicating rows or columns of data on neighboring processors. This capability is also supported through the map interface. If overlap is necessary, it is specified as an additional fourth argument. In Fig. 4, the fourth argument indicates that there is zero overlap between rows and one column overlap between columns. Overlap can be defined for any dimension and does not have to be the same across dimensions.

*C. Processor Indexed Tagged FAmiLy of Line Segments (PITFALLS)*

pPython uses PITFALLS to implement an efficient and general technique for data redistribution. Additionally, Ref. 28 provides an algorithm for determining which pairs of processors need to communicate when redistribution is required and exactly what data needs to be sent.

The pure implementation of PITFALLS has an issue with the block distribution under a certain condition when a dimension is not a multiple of the given number of processors. Fig. 5 demonstrates the short coming of the original PITFALLS implementation. If a dimension of 16 is distributed among the 5 processors with the block distribution, each processor will have 4 elements, resulting in no elements assigned to the last processor. With the enhancement of PITFALLS implementation in pPython, all the processors will get its share with the block distribution and the remainder will be assigned, starting from the rank zero, one-by-one, until everything is distributed.

*D. PythonMPI*

PythonMPI is a pure Python implementation of the most basic MPI functions. The functions used by pPython are `MPI_Init`, `MPI_Comm_size`, `MPI_Comm_rank`, `MPI_Send`, `MPI_Recv`, `MPI_Bcast`, and `MPI_Finalize`. The communication is done through file I/O through a common file system. The characteristics of

PythonMPI (bandwidth, latency performance, lines of code for implementation, and message communication) are what MatlabMPI has already discussed and demonstrated in Ref. 25. In particular, PythonMPI is a very small library, is naturally one-sided messaging (sends can be posted independently of a corresponding receive), and allows for arbitrarily large messages that can be inspected at any time for debugging purposes.

In PythonMPI implementation, all messages are aggregated as a dictionary variable and then saved as a message file using the `h5py` module in a HDF5 format initially. The message files are read recursively to receive the messages by the receiving MPI processes. This process works well for almost all types of messages but the `h5py` utility lacks support for NumPy arrays with complex numbers. To overcome this short-coming with `h5py`, the Python pickle package is finally used to save and load the messages.

Fig. 6 shows both bandwidth and latency results from the pMatlab and pPython implementation. The results are obtained by running the experiments seven times with two MPI processes on a compute node with two Intel Xeon Platinum 8260 2.4 GHz processors (total 48 cores) and 192 Gbytes of RAM memory. The messages are written to and read from a local filesystem with a RAID 0 disk drive. The median values show that pPython has better bandwidth and latency over all message sizes but they are converging to the similar performance at larger messages. This performance comparison indicates that pPython does a lot better job with the underlying file IO and in pPython, pickle performs a little better than h5py.

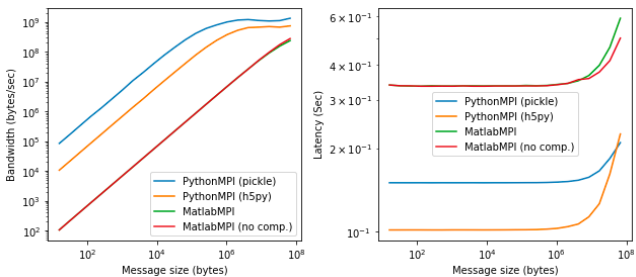

Fig. 6. PythonMPI vs. MatlabMPI. Bandwidth and latency vs. message size. Bandwidth is given as bytes/sec. and latency is given in seconds. The jittery behavior is considered to be a result of high system loads under a production environment.

### E. pPython Parallel Support Functions

A number of pPython parallel support functions are available: `agg`, `global_block_range`, `global_block_ranges`, `grid`, `inmap`, `local`, `put_local`, and `synch`. These functions are used for managing and working with global arrays and have no serial equivalents. These functions allow the user to aggregate data onto one or many processors, to determine which global indices are local to which processors, and to get/put data from/to the local part of a distributed array. These functions, in addition to the `Dmap` class, represent the core of pPython syntax. To support the development process discussed in Section II.A, all of these functions have been implemented to also work on serial Python NumPy arrays and lists so that the pPython code will still work if the pPython maps have been turned off.

### F. HPC Challenge Benchmarks

In this section we focus on benchmark results to determine the performance limits of pPython implementation. We chose to implement a set of standardized benchmarks and compare the performance against those obtained from pMatlab, which has already demonstrated its performance [22] using the HPC Challenge benchmark suite [29]. We have compared the scaling performance by running the benchmarks over a range of processors: 1, 2, 4, 8, 16, 32 on a single node, and 64 on two nodes with 32 on each. For the MPI message communication, the Lustre parallel filesystem running on a DDN 18K storage system is used. The parallel speedup of the pPython programs is the most important attribute as this is what most Python user's seek when make their programs parallel.

#### 1) STREAM

Fig. 7 compares the results obtained from both pPython and pMatlab implementation. The result from the pPython version shows slightly better performance in STREAM for most of the cases as compared to MATLAB when scaling the problem size with Np. The pPython triad bandwidth performance of a single process without multi-threading is about 6.9 Gbytes/sec, which is in line with the result reported in Ref. 30. Both results scale linearly as increasing the number of CPUs up to `Np=8`. STREAM is representative of many of the "pleasingly parallel" applications that many Python users want to implement quickly.

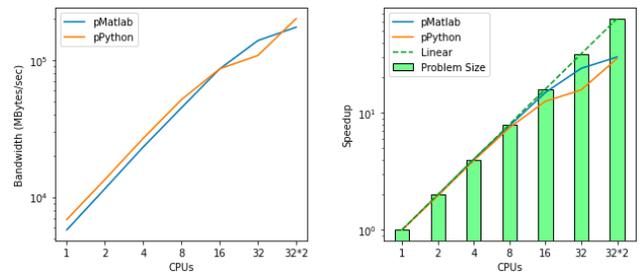

Fig. 7. STREAM throughput and scalability comparison between pPython and pMatlab. Python throughput is on par with that of MATLAB.

#### 2) FFT

The results obtained from both pPython and pMatlab are presented in Fig. 8. The performance of pPython implementation is significantly slower than of pMatlab due to the poor performance of Python FFT method with a single process performance. However, a parallel speedup of 49x at

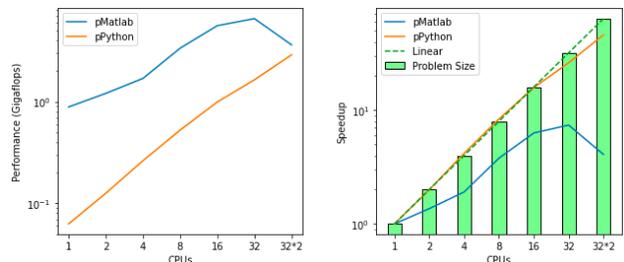

Fig. 8. FFT performance and scalability comparison between pPython and pMatlab. Python performance is much slower than that of MATLAB but scales better.

`Np=64` is achieved, which is significant for a communication heavy benchmark like FFT.

*3) Random Access*

The results obtained from both pPython and pMatlab implementations are presented in Fig. 9. RandomAccess is an extremely communications latency bound benchmark and as expected there is no speedup. Good performance on RandomAccess typically requires advanced direct memory access implementations in a low-level language on low-latency networks.

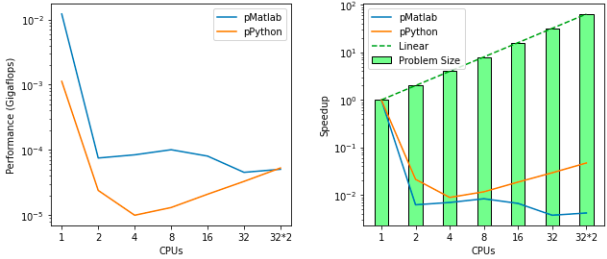

Fig. 9. RandomAccess performance and scalability comparison between pPython and pMatlab. Python performance is slower than that of MATLAB except Np=64.

*4) High Performance Linpack (Top500)*

The results from both pMatlab and pPython versions are presented in Fig. 10. We have found that the LU factorization method (`scipy.linalg.lu`) used in pPython is aggressively multithreaded using the entire resources on the node. Thus, it is important to disable multithreading when more than one pPython processes are running on the same node. We are able to scale pPython up to `Np=32` within a node without out-of-memory error and scale out to two nodes with `Np=64` whereas pMatlab implementation encountered an indexing error when scale out to two nodes at `Np=64` for the same problem size. For a single proces pPyton HPL benchmark achieved about 12.5 Gigaflops with a matrix size of 4K elements in each direction. Considering the theoretical peak performance of 76.8 Gigaflops per core, pPython implementation needs some more work to improve the performance.

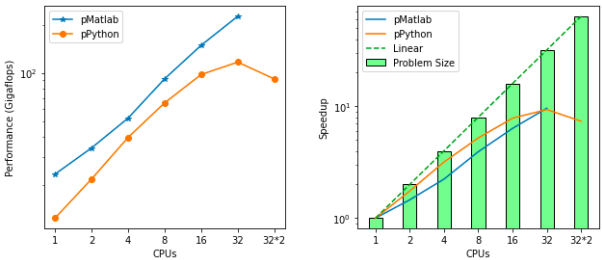

Fig. 10. HPL/Top500 benchmark performance and scalability comparison between pPython and pMatlab. Python performance is comparable to that of MATLAB.

IV. ADDITIONAL OBSERVATIONS

While implementing pPython on a Python environment, we found that the majority of the code was easily implemented in Python because Python is high-productive scripting languages along with a large number of scientific and engineering packages. However, there are also some nuances that can cause some unexpected behavior. As such, we would like to document some of these observations and remedies that are implemented in the pPython package and may be of use to others implementing parallel libraries in Python.

*A. Ordering*

When you create a 2 or higher dimensional array, how you lay out the array makes a significant difference in the final array. MATLAB stores arrays in column-major order (FORTRAN-style) and Python stores arrays in row-major order (C-style). When a map is defined with 2-D dimensional processor grid, the processor ID (equivalent to a MPI rank) is stored in column-major order in pMatlab as shown in Fig. 1. We have added the keyword, 'order', in order to match with the pMatlab behavior when creating a map object if needed for converting a pMatlab code to a pPython code. Also, it is important to note that, when flattening or reshaping an array, the ordering makes a significant difference. Thus, when converting a pMatlab code to a pPython code, careful attention should be taken to the ordering of array elements.

*B. Randomness*

When running a pMatlab code with multiple processes, each of the processes generates exactly the same random numbers unless the random seed is defined differently on each process. However, with pPython, each process generates completely different random numbers by default.

*C. Indexing and Range Expression*

Array indexing is different between MATLAB and Python. MATLAB index starts from 1 but Python index starts from 0. Also, the following expression in MATLAB extracts a 2x2 submatrix: `A(2:3,4:5)`, where starting from $2^{nd}$ row to $3^{rd}$ row and from $4^{th}$ column to $5^{th}$ column. However, the equivalent expression in Python is `A[1:3,3:5]` because of differences in the indexing and range expression. This is one of the most common differences when porting pMatlab to pPython.

*D. Subscripted Reference and Assignment*

pMatlab distributed array operations such as addition, subtraction, multiplication on submatrix is implemented with subscripted reference and assignment operations with the operator overloading. However, in pPython, distributed array operations are implemented in the distributed array class with the Python intrinsic functions such as the __add__, __sub__, and __mul__ methods. The subscripted assignment operation is implemented with the __setitem__ method in the distributed array class.

V. SUMMARY

In this paper, we have presented pPython, which enables a parallel programming to achieve good speed up without sacrificing the ease of programming in Python by implementing partitioned global array semantics (PGAS) with porting pMatlab, MatlabMPI and gridMatlab in Python. Some preliminary results of HPC Challenge benchmark suit are compared with those obtained from pMatlab implementation. It should be noted that, during the conversion, we have focused more on the functional correctness than anything else. Current pPython performance shows better performance in MPI message communication but is slower than that of pMatlab with most of HPC Challenge benchmark examples. Initial pPython implementation experiences significant memory requirements and fails to handle large problem sizes. We have found the main cause of large memory usage is attributed to inefficient use of array indexing. After addressing the issue, pPython implementation can handle the problem sizes equivalent to or greater than what pMatlab can handle.

We have also discussed some of the challenges while converting pMaltab into pPython. There are many more issues that are not covered here due to the length limitation of the paper but we can address them when we publish the source code. We still have a number of bugs to address before releasing the code. Performance and robustness are two major topics to be focused on in the near term.

Overall pPython provides a simple and maintainable way to easily make Python programs run in parallel with reasonable performance scalability.

ACKNOWLEDGMENTS

The authors wish to acknowledge the following individuals for their contributions and support: Bob Bond, Alan Edelman, Jeff Gottschalk, Charles Leiserson, Joseph McDonald, Steve Rejto, Matthew Weiss, and Marc Zissman.